\documentclass{article}
\usepackage{amsmath,amssymb,cite,epsfig,graphicx,fancyhdr}
\begin{document}
\lhead{Complex solutions to $P_{IV}$ through SUSY QM}
\rhead{D. Bermudez, D. J. Fern\'andez C.}

\title{Complex solutions to Painlev\'e IV equation \\ through supersymmetric quantum mechanics}
\author{David Berm\'udez\footnote{{\it email:} dbermudez@fis.cinvestav.mx}\,  
and David J. Fern\'andez C.\footnote{{\it email:} david@fis.cinvestav.mx} \\
{\sl Departamento de F\'{\i}sica, Cinvestav, A.P. 14-740, 07000 M\'exico D.F., Mexico}}

\date{}

\maketitle

\begin{abstract}
In this work, supersymmetric quantum mechanics will be used to obtain complex solutions to Painlev\'e IV equation with real parameters. We will also focus on the properties of the associated Hamiltonians, i.e. the algebraic structure, the eigenfunctions and the energy spectra.\\

\end{abstract}

\section{Introduction}

In recent years a growing interest in studying non-linear models has appeared, since they can be used to describe many physical situations. This leads typically to the analysis of non-linear differential equations, which are harder to solve than the linear ones. However, there are solution methods to solve them, which mainly consist in reducing non-linear equations to linear ones. This approach has been employed for obtaining solutions of the Painlev\'e equations, which we are particularly interested in. Although discovered from strictly mathematical considerations, the Painlev\'e equations are widely used nowadays to describe physical phenomena \cite{AC92}.

When is based on supersymmetric quantum mechanics (SUSY QM), this method transforms the non-linear equations into Schr\"odinger type ones. In addition, if we end up with a previously solved Schr\"odinger equation this problem will be further simplified. In particular, this works for the Painlev\'e IV equation ($P_{IV}$) \cite{VS93,Adl94}, which is relevant in fluid mechanics, non-linear optics, quantum gravity \cite{Win92}, and SUSY QM applied to the harmonic oscillator \cite{BF11,BF11b}. In fact, a simple method to generate solutions of $P_{IV}$ can be supplied, since the SUSY technique provides explicit expressions for the relevant extremal states and formulae connecting them with the $P_{IV}$ solutions \cite{CFNN04,MN08}.

This contribution is organized as follows: in Sections 2 and 3 we shall present the general framework of SUSY QM and polynomial Heisenberg algebras (PHA). In Section 4 we will generate the complex solutions to $P_{IV}$, while in Section 5 we will study the eigenfunctions and eigenvalues of the non-hermitian Hamiltonians. Our conclusions shall be presented in Section 6.

\section{Higher-order SUSY QM}

In the $k$-th order SUSY QM one starts from a given solvable Hamiltonian
\begin{equation}
H_0  = -\frac12 \frac{d^2}{d x^2} + V_0(x),
\end{equation}
and generates a chain of standard (first-order) intertwining relations \cite{AIS93,Fer10}
\begin{eqnarray}
&& \hskip2.3cm H_j A_j^{+}  = A_j^{+} H_{j-1}, \quad  H_{j-1}A_j^{-} = A_j^{-}H_j, \\
&& H_j = -\frac12 \frac{d^2}{d x^2} + V_j(x), \quad
A_j^{\pm} = \frac{1}{\sqrt{2}}\left[\mp \frac{d}{d x} + \alpha_j(x,\epsilon_j)\right], \quad j = 1,\dots,k.
\end{eqnarray}
Hence, the following equations must be satisfied
\begin{align}
& \alpha_j'(x,\epsilon_j) + \alpha_j^2(x,\epsilon_j) = 2[V_{j-1}(x) - \epsilon_j],  \label{rei} \\ 
& \hskip1.0cm  
V_{j}(x) = V_{j-1}(x) - \alpha_j'(x,\epsilon_j). 
\end{align}
We are interested in the final Riccati solution $\alpha_{k}(x,\epsilon_{k})$, which turns out to be determined either by $k$ solutions $\alpha_1(x,\epsilon_j)$ of the initial Riccati equation or by $k$ solutions $u_j$ of the Schr\"odinger equation obtained through the change $\alpha_1(x,\epsilon_j) = u_j'/u_j$:
\begin{equation}
H_0 u_j = - \frac12 u_j'' + V_0(x)u_j = \epsilon_j u_j, \quad j=1,\dots,k. \label{usch}
\end{equation}

\section{General framework of PHA}

A $m$-th order PHA is a deformation of the Heisenberg-Weyl algebra of kind \cite{CFNN04,FH99}:
\begin{align}
[H,L^\pm] 	& = \pm L^\pm , \qquad
[L^-,L^+] 	\equiv Q_{m+1}(H+1) - Q_{m+1}(H) = P_m(H) , \\ 
& \hskip2.0cm 
Q_{m+1}(H) 	= L^+ L^- = \prod_{i=1}^{m+1} \left(H - \mathcal{E}_i\right),
\end{align}
where $Q_{m+1}(x)$ is a $(m+1)$-th order polynomial in $x$, $P_m(x)$ is a polynomial of order $m$ in $x$ and $\mathcal{E}_i$ are the zeros of $Q_{m+1}(H)$, which correspond to the energies associated to the extremal states of $H$.

For the second-order PHA ($m=2$), let us suppose from now on that $L^+$ is a third-order differential ladder operator of the form:
\begin{align}
L^+  = L_1^+ L_2^+ , \qquad
L_1^+ = \frac{1}{\sqrt{2}}\left[-\frac{d}{d x} + f(x) \right], \qquad
L_2^+ = \frac12\left[ \frac{d^2}{d x^2} + g(x)\frac{d}{d x} +
h(x)\right].
\end{align}
Using the standard first and second-order SUSY QM and decoupling the resulting system gives rise to equations for $f$, $h$, and $V$ in terms of $g$ and the following condition:
\begin{equation}
g'' = \frac{g'^2}{2g} + \frac{3}{2} g^3 + 4xg^2 + 2\left(x^2 - a \right) g + \frac{b}{g}.
\end{equation}
This is the Painlev\'e IV equation ($P_{IV}$) with parameters $a =\mathcal{E}_2 + \mathcal{E}_3-2\mathcal{E}_1 -1$ and $b = - 2(\mathcal{E}_2 - \mathcal{E}_3)^2$. Hence, if the three quantities $\mathcal{E}_i$ are real, we will obtain real parameters $a,b$ for the corresponding $P_{IV}$.

The key point now comes from the inverse method: instead of solving directly $P_{IV}$ to get a system ruled by a second-order PHA let us choose a system having this algebra (i.e. with third-order ladder operators) in order to obtain then the solutions to $P_{IV}$.

\section{Complex solutions to $P_{IV}$ with real parameters}

Let us use a theorem stating the conditions for the hermitian higher-order SUSY partners of the harmonic oscillator Hamiltonian to have second-order PHA (see \cite{BF11}). The main requirement is that the $k$ Schr\"odinger seed solutions have to be connected in the way
\begin{align}
u_j=(a^{-})^{j-1}u_1,& \quad \label{us}\ \quad
\epsilon_j=\epsilon_1-(j-1),  \quad j=1,\dots , k,
\end{align}
where $a^{-}$ is the standard annihilation operator, the only free seed $u_1$ is a nodeless real solution of Eq.~\eqref{usch} associated to a real factorization energy $\epsilon_1$ such that $\epsilon_1<E_0=1/2$.

In order to overcome the last restriction, so that $\epsilon_1 > E_0$ but the SUSY transformation being still non-singular, we need to use complex seed solutions. Let us choose them as complex linear combinations of the two standard linearly independent real solutions so that, up to an unessential factor, we will employ \cite{ACDI99}:
\begin{equation}
u(x;\epsilon ) = e^{-x^2/2}\left[ {}_1F_1\left(\frac{1-2\epsilon}{4},\frac12;x^2\right)
 + x(\lambda + i\kappa)\, {}_1F_1\left(\frac{3-2\epsilon}{4},\frac32;x^2\right)\right], \label{u1}
\end{equation}
where $\lambda, \kappa \in \mathbb{R}$ and $_1F_1$ is the confluent hypergeometric (Kummer) function.

\begin{figure}
\includegraphics[scale=0.35]{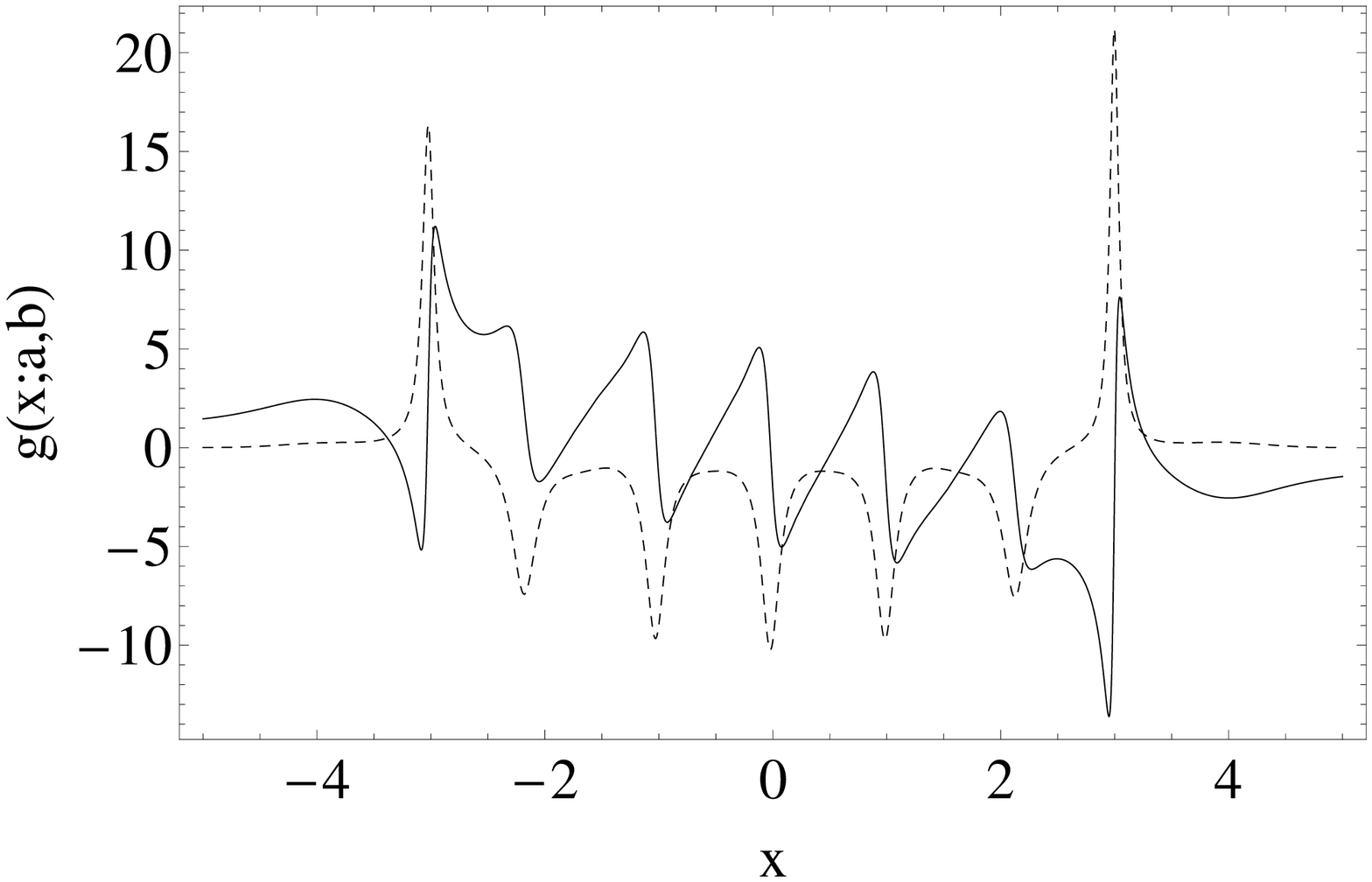}\hspace{5mm}
\includegraphics[scale=0.345]{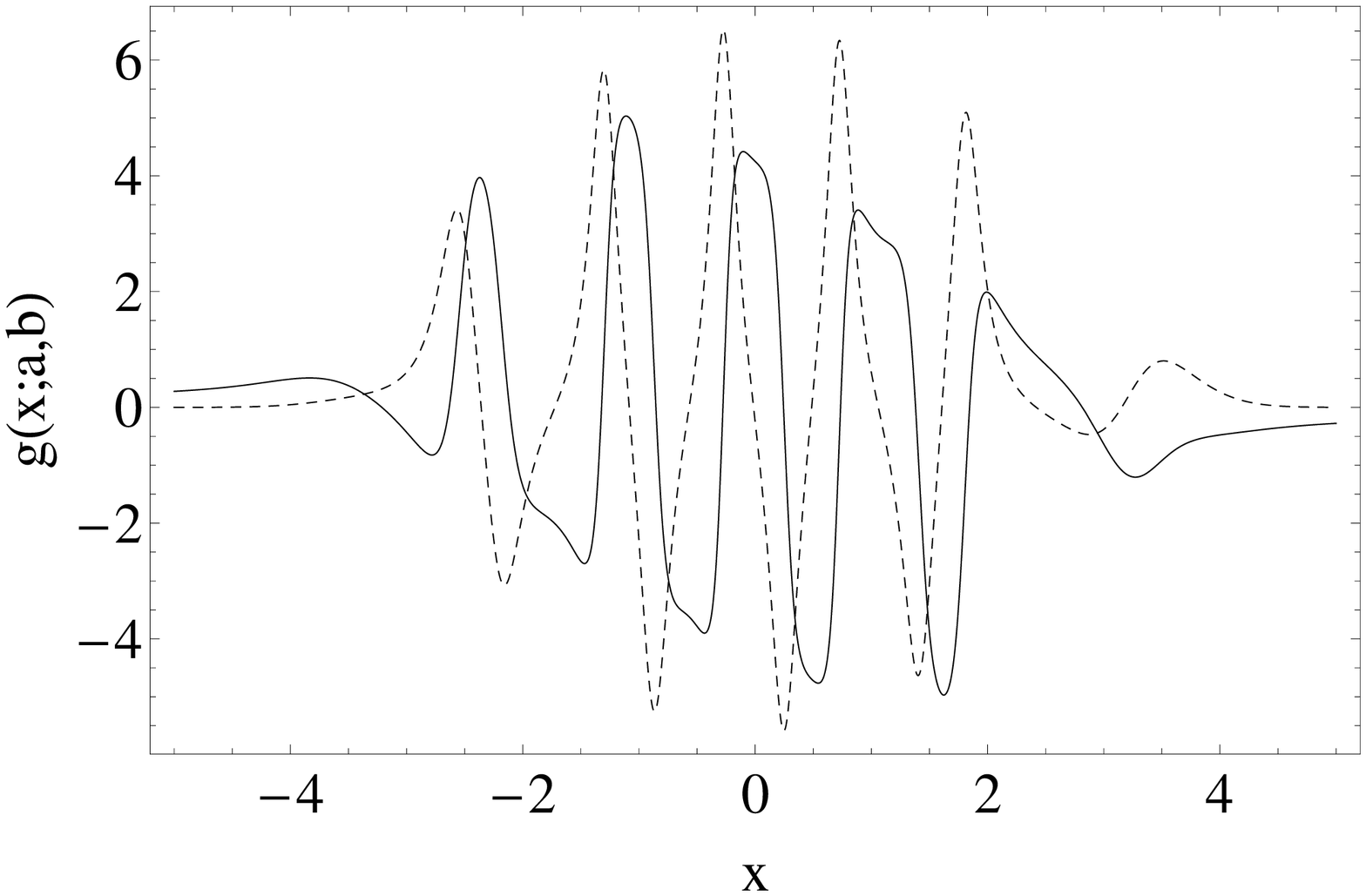}
\vspace{-5mm}
\caption{Real (solid curve) and imaginary (dashed curve) parts of some complex solutions to $P_{IV}$. The left plot corresponds to $a=-5/2$, $b=-121/2$ ($k=2$, $\epsilon_1=5$, $\lambda=1$, $\kappa=5$) and the right one to $a=9$, $b_3=-2$ ($k=1$, $\epsilon_1=5$, $\lambda=\kappa=1$).} \label{gcomplex}
\end{figure}

Hence, through this formalism we will obtain simultaneously the $k$-th order SUSY partner potential $V_k(x)$ of the harmonic oscillator and the corresponding $P_{IV}$ solution $g(x;\epsilon_1)$, both of which are complex, in the following way
\begin{align}
V_k(x) & = x^2/2 - \{\ln [W(u_1,\dots,u_k)]\}'' , \qquad
g(x;\epsilon_1) = - x - \{\ln[\psi_{\mathcal{E}_1}(x)]\}', \label{solg}
\end{align}
which can be compared with those obtained in \cite{BCH95} through B\"acklund transformations. Note that the extremal states of $H_{k}$ and their corresponding energies are given by
\begin{eqnarray}
&& \psi_{\mathcal{E}_1} \propto \frac{W(u_1,\dots,u_{k-1})}{W(u_1,\dots,u_k)},
\quad \psi_{\mathcal{E}_2} \propto B_k^{+} e^{-x^2/2},
\quad \psi_{\mathcal{E}_3} \propto B_k^{+} a^{+} u_1, \\
&&  \hskip0.2cm \mathcal{E}_1 = \epsilon_k = \epsilon_1 - (k - 1), \quad \hskip0.1cm \mathcal{E}_2 = \frac{1}{2}, \qquad \qquad \hskip0.2cm 
 \mathcal{E}_3 = \epsilon_1 + 1,
\end{eqnarray}
where $B_k^{+}=A_k^{+}\dots A_1^{+}$. The real and imaginary parts of the complex solutions $g(x;a,b)$ for two particular choices of real parameters $a,b$ are plotted in Fig.~\ref{gcomplex} .

\section{Non-hermitian Hamiltonians}

Let us analyze the Hamiltonian $H_k$ obtained by the complex SUSY transformation. The real case with $\epsilon_1<1/2$, leading to hermitian Hamiltonians, has been studied previously \cite{AIS93,Mie84}. The spectrum $\text{Sp}(H_k)$ consists of an infinite ladder plus a finite one: there are two extremal states (both annihilated by $L^{-}$) from which the two ladders start, one associated to $\epsilon_k$ and the other to $E_0=1/2$; since the ladder starting from $\epsilon_k$ ends at $\epsilon_1$, the eigenfunction associated to $\epsilon_1$ is annihilated by $L^{+}$. The actions of $L^{\pm}$ onto any other eigenstate of $H_k$ are non-null and connect only eigenstates belonging to the same ladder.

In the complex case, the new Hamiltonians necessarily have complex eigenfunctions (square-integrable), although the associated eigenvalues are still real. The resulting spectra for the non-hermitian Hamiltonians $H_k$ obey the same criteria as the real case, namely, they are composed of an infinite ladder plus a finite one which now could be placed, either fully or partially, above $E_0=1/2$.

\begin{figure}
\begin{center}
\includegraphics[scale=0.1]{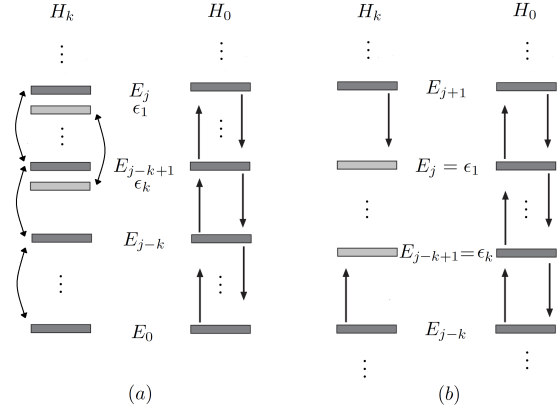}
\end{center}
\vspace{-5mm}
\caption{Spectra of $H_k$ and $H_0$ for $\epsilon_1>1/2$ and (a) $\epsilon_1\neq E_j$; (b) $\epsilon_1=E_j$. In (a), Sp($H_k$) still contains one finite and one infinite ladders, but this is not true for (b). The dark bars represent the original and mapped eigenstates of $H_0$ and $H_k$, while the light ones do for the $k$ new levels of $H_k$.} \label{1susy}
\end{figure}

The case when $\epsilon_1$ belongs to the spectrum of the original harmonic oscillator Hamiltonian is worth of a detailed study. Thus, let us consider a $k$-order SUSY transformation with $\epsilon_1=E_j$, $j>k$, and $u_1$ given by Eq.~\eqref{u1}, i.e., $u_1$ is a complex linear combination of the eigenfunction $\psi_j$ of $H_0$ and the other linearly independent solution of the Schr\"odinger equation. It is straightforward to see that the action of the third-order ladder operators applied to the eigenstate with energy $E_{l}$ is given by
\begin{align}
L^{-}(B_k^{+}\psi_{l})\propto B_k^{+}\psi_{l-1},
& \text{ for } l\not\in \{0,j-k+1\};
& L^{-}(B_k^{+}\psi_{0})=L^{-}(B_k^{+}\psi_{j-k+1})=0,\\
L^{+}(B_k^{+}\psi_{l}) \propto B_k^{+}\psi_{l+1},
& \text{ for } l\neq j;
& L^{+}(B_k^{+}\psi_{j}) =0.\phantom{...........................}
\end{align}
These results do not match with the established criteria for the non-singular real and complex cases with $\epsilon_1 \neq E_j$ since now it turns out that:
\begin{align}
E_{j+1},&\quad L^{-}(B_k^{+}\psi_{j+1})  \propto B_k^{+}\psi_{j} \neq 0,\\
E_{j-k},&\quad L^{+}(B_k^{+}\psi_{j-k})   \propto B_k^{+}\psi_{j-k+1} \neq 0,
\end{align}
where the shown energies correspond to the departure state. The resulting Hamiltonian is isospectral to the harmonic oscillator but with a special algebraic structure since now two states (those associated to $E_{j+1}$ and $E_{j-k}$) are connected just in one way with the adjacent ones (see a diagram representing this in Fig.~\ref{1susy}).

\section{Conclusions}

We have described a simple method to obtain real and complex solutions $g(x;a,b)$ of the $P_{IV}$ with real parameters $a,b$. We have studied the properties of the resulting solutions, and we have analyzed the algebras, the eigenfunctions and the spectra of the non-hermitian SUSY generated Hamiltonians. Further extensions of this technique to obtain $P_{IV}$ solutions associated to complex parameters $a,b$ are under study.


\section*{Acknowledgement}
The authors acknowledge the financial support of Conacyt (M\'exico) through project 152574. DB also acknowledges the Conacyt PhD scholarship 219665.


\begin{thebibliography}{99}

\bibitem{AC92} M.J. Ablowitz, P.A. Clarkson. {\it Solitons, nonlinear evolution equations and inverse scattering}, Cambridge University Press, New York, 1992.

\bibitem{VS93} A.P. Veselov, A.B. Shabat,
{\it Funct. Anal. Appl.} {\bf 27} (1993) 81-96.

\bibitem{Adl94} V.E. Adler,
{\it Physica D} {\bf 73} (1994) 335-351.

\bibitem{Win92} P. Winternitz,
in {\it Painlev\'e trascendents, their asymptotics and physical applications}, NATO ASI Series B, New York (1992) 425-431.

\bibitem{BF11} D. Berm\'udez, D.J. Fern\'andez,
{\it SIGMA} {\bf 7} (2011) 025, 14 pages.

\bibitem{BF11b} D. Berm\'udez, D.J. Fern\'andez,
{\it Phys. Lett. A} {\bf 375} (2011) 2974-2978.

\bibitem{CFNN04} J.M. Carballo, D.J. Fern\'andez, J. Negro, L.M. Nieto,
{\it J. Phys. A: Math. Gen.} 37 (2004) 10349-10362.

\bibitem{MN08} J. Mateo, J. Negro,
{\it J. Phys. A: Math. Theor.} {\bf 41} (2008) 045204, 28 pages.

\bibitem{AIS93} A.A. Andrianov, M. Ioffe, V. Spiridonov,
{\it Phys. Lett. A} {\bf 174} (1993) 273-279.

\bibitem{Fer10} D.J. Fern\'andez,
{\it AIP Conf. Proc.} {\bf 1287} (2010) 3-36.

\bibitem{FH99} D.J. Fern\'andez, V. Hussin,
{\it J. Phys. A: Math. Gen.} {\bf 32} (1999) 3603-3619.

\bibitem{ACDI99} A.A. Andrianov, M.V. Ioffe, F. Cannata, J.P. Dedonder,
{\it Int. J. Mod. Phys. A} {\bf 14} (1999) 2675-2688.

\bibitem{BCH95} A.P. Bassom, P.A. Clarkson, A.C. Hicks,
{\it Stud. Appl. Math.} {\bf 95} (1995) 1-75.

\bibitem{Mie84} B. Mielnik,
{\it J. Math. Phys.} {\bf 25} (1984) 3387-3389.

\end{thebibliography}
\end{document}